\documentclass[prl,twocolumn,longbibliography,
superscriptaddress,
 amsmath,amssymb,
 aps,floatfix
]{revtex4-1}
\usepackage{gensymb}
\usepackage{graphicx}
\usepackage{dcolumn}
\usepackage{bm}
\usepackage{multirow}
\usepackage{textcomp, gensymb}
\usepackage{
}

\newcommand{\cgt}{Cr$_2$Ge$_2$Te$_6$}

\newcommand{\tc}{T$_C$}

\begin{document}
\title{Two-Step Electronic Response to Magnetic Ordering in a van der Waals Ferromagnet}

\author{Han Wu}
\affiliation{Department of Physics and Astronomy and Rice Center for Quantum Materials, Rice University, Houston, TX, 77005 USA}

\author{Jian-Xin Zhu}
\affiliation{Theoretical Division, Los Alamos National Laboratory, Los Alamos, New Mexico 87545, USA}
\affiliation{Center for Integrated Nanotechnologies, Los Alamos National Laboratory, Los Alamos, New Mexico 87545, USA}

\author{Lebing Chen}
\affiliation{Department of Physics, University of California at Berkeley, Berkeley, California 94720, USA}
\affiliation{Department of Physics and Astronomy and Rice Center for Quantum Materials, Rice University, Houston, TX, 77005 USA}

\author{Matthew W Butcher}
\affiliation{Department of Physics and Astronomy and Rice Center for Quantum Materials, Rice University, Houston, TX, 77005 USA}

\author{Ziqin Yue}
\affiliation{Department of Physics and Astronomy and Rice Center for Quantum Materials, Rice University, Houston, TX, 77005 USA}
\affiliation{Applied Physics Graduate Program, Smalley-Curl Institute, Rice University, Houston, Texas, 77005, USA}

\author{Dongsheng Yuan}
\affiliation{Materials Sciences Division, Lawrence Berkeley National Laboratory, Berkeley, California 94720, USA}
\affiliation{National Institute for Materials Science (NIMS), 1-1 Namiki, Tsukuba, Ibaraki 305-0044, Japan}

\author{Yu He}
\affiliation{Department of Applied Physics, Yale University, New Haven, CT 06511}

\author{Ji Seop Oh}
\affiliation{Department of Physics and Astronomy and Rice Center for Quantum Materials, Rice University, Houston, TX, 77005 USA}
\affiliation{Department of Physics, University of California at Berkeley, Berkeley, California 94720, USA}

\author{Jianwei Huang}
\affiliation{Department of Physics and Astronomy and Rice Center for Quantum Materials, Rice University, Houston, TX, 77005 USA}

\author{Shan Wu}
\affiliation{Department of Physics, University of California at Berkeley, Berkeley, California 94720, USA}

\author{Cheng Gong}
\affiliation{Department of Electrical and Computer Engineering and Quantum Technology Center, University of Maryland, College Park, Maryland 20742, USA}

\author{Yucheng Guo}
\affiliation{Department of Physics and Astronomy and Rice Center for Quantum Materials, Rice University, Houston, TX, 77005 USA}

\author{Sung-Kwan Mo}
\affiliation{Advanced Light Source, Lawrence Berkeley National Laboratory, Berkeley, CA 94720, USA}

\author{Jonathan D. Denlinger}
\affiliation{Advanced Light Source, Lawrence Berkeley National Laboratory, Berkeley, CA 94720, USA}

\author{Donghui Lu}
\affiliation{Stanford Synchrotron Radiation Lightsource, SLAC National Accelerator Laboratory, Menlo Park, California 94025, USA}

\author{Makoto Hashimoto}
\affiliation{Stanford Synchrotron Radiation Lightsource, SLAC National Accelerator Laboratory, Menlo Park, California 94025, USA}

\author{Matthew B. Stone}
\affiliation{Neutron Scattering Division, Oak Ridge National Laboratory, Oak Ridge, Tennessee 37831, USA}

\author{Alexander I. Kolesnikov}
\affiliation{Neutron Scattering Division, Oak Ridge National Laboratory, Oak Ridge, Tennessee 37831, USA}

\author{Songxue Chi}
\affiliation{Neutron Scattering Division, Oak Ridge National Laboratory, Oak Ridge, Tennessee 37831, USA}

\author{Junichiro Kono}
\affiliation{Department of Electrical and Computer Engineering, Rice University, Houston, Texas 77005, USA}
\affiliation{Smalley-Curl Institute, Rice University, Houston, Texas, 77005, USA}
\affiliation{Department of Physics and Astronomy and Rice Center for Quantum Materials, Rice University, Houston, TX, 77005 USA}
\affiliation{Department of Material Science and NanoEngineering, Rice University, Houston, Texas 77005, USA
}


\author{Andriy H. Nevidomskyy}
\affiliation{Department of Physics and Astronomy and Rice Center for Quantum Materials, Rice University, Houston, TX, 77005 USA}

\author{Robert J. Birgeneau}
\affiliation{Department of Physics, University of California at Berkeley, Berkeley, California 94720, USA}
\affiliation{Materials Sciences Division, Lawrence Berkeley National Laboratory, Berkeley, California 94720, USA}
\affiliation{Department of Materials Science and Engineering, University of California, Berkeley, USA}

\author{Pengcheng Dai}
\affiliation{Department of Physics and Astronomy and Rice Center for Quantum Materials, Rice University, Houston, TX, 77005 USA}

\author{Ming Yi}
\email{mingyi@rice.edu}
\affiliation{Department of Physics and Astronomy and Rice Center for Quantum Materials, Rice University, Houston, TX, 77005 USA}

\date{\today}%

\begin{abstract}
The two-dimensional (2D) material Cr$_2$Ge$_2$Te$_6$ is a member of the class of insulating van der Waals magnets.
Here, using high resolution angle-resolved photoemission spectroscopy in a detailed temperature dependence study, we identify a clear response of the electronic structure to a dimensional crossover in the form of two distinct temperature scales marking onsets of modifications in the electronic structure. Specifically, we observe Te $p$-orbital-dominated bands to undergo changes at the Curie transition temperature T$_C$ while the Cr $d$-orbital-dominated bands begin evolving at a higher temperature scale. Combined with neutron scattering, density functional theory calculations, and Monte Carlo simulations, we find that the electronic system can be consistently understood to respond sequentially to the distinct temperatures at which in-plane and out-of-plane spin correlations exceed a characteristic length scale. Our findings reveal the sensitivity of the orbital-selective electronic structure for probing the dynamical evolution of local moment correlations in vdW insulating magnets.

\end{abstract}

\maketitle

Exploring the magnetism in quasi-2D materials has been a fascinating subject in quantum physics for more than five decades. This field has received strong stimuli from both the discovery of high temperature superconductivity in the lamellar copper oxides in the late 1980s and, more recently, by studies of the ferromagnetism in van der Waals (vdW) materials~\cite{Bob_lamellar,Gong2017,Huang2017,Mak2019,Burch2018,Cheng2019,Tokura2017,Huang2020,Wang2020,Castro2009}. 
The chromium tellurides Cr$_2$X$_2$Te$_6$ (X=Ge, Si and Sn) belong to a category of insulators with intrinsic long-range ferromagnetic order down to the 2D regime~\cite{Zhang2019,Siberchicot1996,Jiang2020,Li2018,Watson2020,Cheng2019,Tokura2017,Huang2020,Wang2020,Castro2009,Allen2010,Miao2021,Corasaniti2022-add1,Corasaniti2020-add2,Corasaniti2021-add3,Yang2022-add4,Ghosh2023-add5,Menichetti2019-add6,Menichetti2023-add7,Zhu2016-add8,Huang2021-my5}. Cr$_2$Ge$_2$Te$_6$, in particular, exhibits ferromagnetism with a \tc~that ranges from 65 K in bulk to around 40 K when exfoliated down to bilayer flakes, with the easy axis along the $c$ direction~\cite{Zuti2004,Liu2016,MacDonald2005,Geim2013,liu2017,Suzuki2019,Yilmaz_2021}. 
The nature of the magnetism in two dimensions in the vdW magnets can be understood to originate from the magnetic anisotropy that can counteract the strong thermal fluctuations. A neutron scattering study on Cr$_2$Si$_2$Te$_6$ (\tc = 35 K) has provided direct evidence on the development of the magnetic order, where the exchange interaction along the c direction is much smaller than that in the in-plane directions, and the dynamic correlations can persist in the $ab$ plane up to at least 300 K~\cite{williams2015}. Such an effective 3D to 2D dimensional crossover behavior of the magnetic order in Cr$_2$Si$_2$Te$_6$ has also been confirmed by spin correlation driven lattice distortions~\cite{Ron2019}. Electronically, these Cr-based vdW ferromagnets are gapped at the Fermi level due to strong Coulomb repulsion~\cite{Zhang2019, Watson2020,Wu2022-my4}. In contrast to the metallic Fe$_n$GeTe$_2$ (n=3-5) systems, Cr$_2$Ge$_2$Te$_6$ as an insulating magnet with a simple magnetic order untangled with other competing or intertwined electronic orders is an ideal platform to study the impact of low dimensional magnetism on the electronic degree of freedom~\cite{Zhu2016-add8,Wu2023-my7,Wu2023-my6}.

\begin{figure*}
\includegraphics[width=0.95\textwidth]{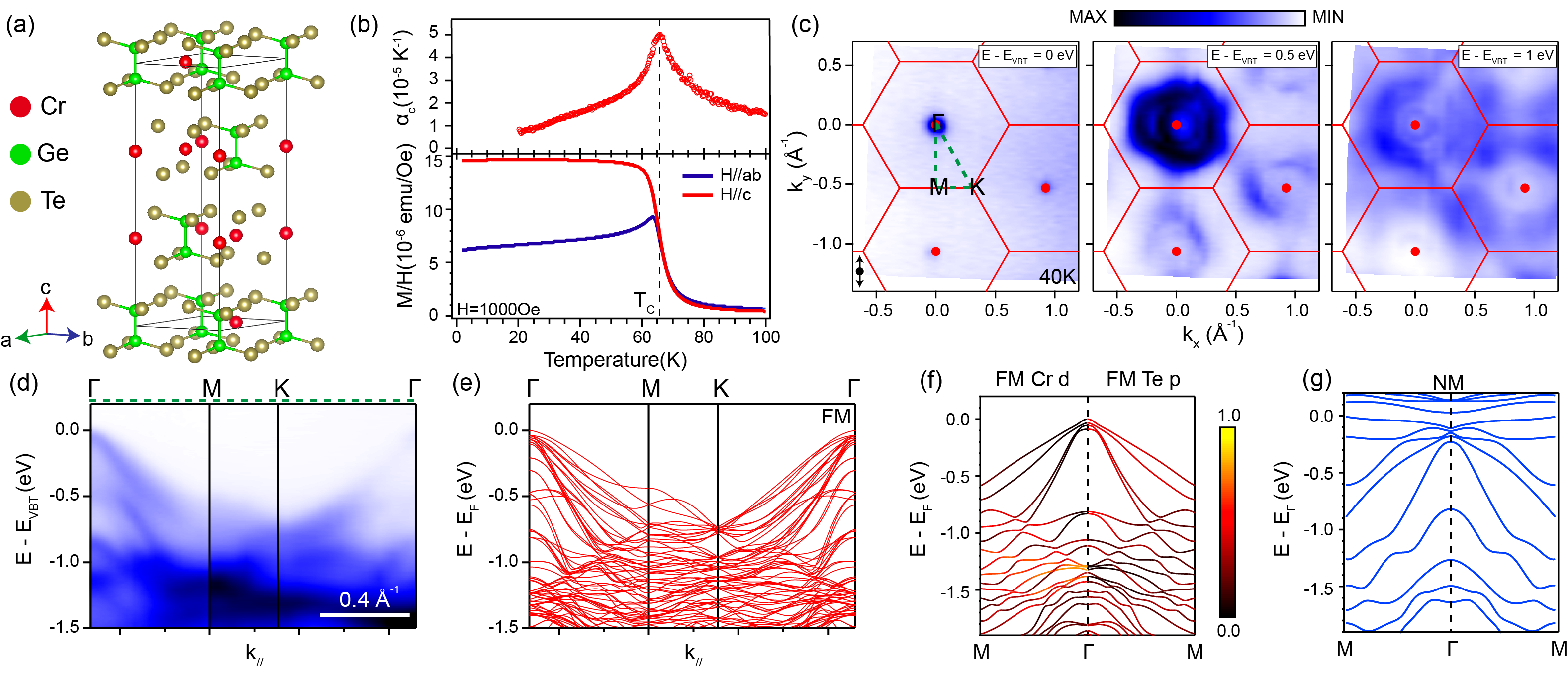}
\caption{Crystal structure, magnetization, and electronic structure of Cr$_2$Ge$_2$Te$_6$.
(a) Crystal structure. (b) Dilatometry measurement  of the c-axis thermal coefficient $\alpha$ (upper) and magnetization (lower). (c) ARPES constant energy contours measured in the FM phase at 40 K, with energy referenced to the valence band top (VBT). BZ centers and boundaries are shown, along with the polarization vector. (d) Spectral image along $\Gamma$-M-K-$\Gamma$ as marked in (c). (e) DFT ferromagnetic calculation along $\Gamma$-M-K-$\Gamma$. (f) DFT calculated band structures of the ferromagnetic Cr$_2$Ge$_2$Te$_6$ projected onto the Cr-d and Te-p orbitals. (g) DFT calculations for the non-magnetic state without local moments.
}\label{fig:Fig1}
\end{figure*}

Here we report the observation of the electronic response to the development of the spin correlations across a wide range of temperatures in the vdW magnet Cr$_2$Ge$_2$Te$_6$ via angle-resolved photoemission spectroscopy (ARPES). By mapping out the temperature dependent band structure and the one-electron spectral evolution across \tc, we observe two types of band evolutions. One group associated with Cr d orbitals that exhibits a gradual shift with an onset temperature well above \tc, and another associated with Te p orbitals that rapidly shift near T$_C$. 
From a combination of neutron scattering, Monte Carlo simulations, and Density Functional Theory (DFT) calculations, we arrive at a holistic understanding of the sequential electronic response as tracking the development of in-plane and out-of-plane spin correlations. Due to the anisotropy in the in-plane and out-of-plane exchange couplings, the in-plane correlation length exceeds that of the lattice constant at a temperature roughly twice that of \tc, while the out-of-plane spin correlation length reaches that of a lattice constant much closer to \tc, affecting more significantly the Te $p$ orbitals through the Cr-Te-Cr superexchange interactions near \tc. Our results provide a consistent understanding of the two-step evolution of the electronic response to the interplay between local moments in 2D magnets, and demonstrate the sensitivity of using orbital-dependent electronic structure to track evolution of spin correlations in these vdW magnets.

Cr$_2$Ge$_2$Te$_6$ forms in the space group 148 ($\bar{R}$3) in a layered structure with weak vdW coupling between adjacent layers (Fig.~\ref{fig:Fig1}(a)). The lattice parameters at 15K determined from neutron scattering are $a$ = 6.832~\AA ~and $c$ = 20.386~\AA~, consistent with previous reports~\cite{Li2018,Watson2020,Wang2023}. The magnetic anisotropy favors the easy axis to be along the c direction. As shown in Fig. 1(b), our field-cooled magnetization measurements show a clear paramagnetic (PM) to ferromagnetic (FM) order transition at 65 K, in agreement with previous studies~\cite{Li2018,Watson2020,Suzuki2019,Yilmaz_2021}. In addition, the PM to FM transition can also be clearly observed in our dilatometry measurement of the c-axis thermal expansion coefficient, $\alpha$ (Fig.~\ref{fig:Fig1}b).


Next, we present the ARPES measured electronic structure in the FM phase. The electronic structure of Cr$_2$Ge$_2$Te$_6$ mimics that of a semiconductor, with hole-like bands at the $\Gamma$ points of the Brillouin zone (BZ) as the valence band top (VBT), consistent with previous reports~\cite{Li2018}. We reference the energy axis to the VBT. From the series of constant energy contours, the electronic structure of the valence bands can be seen to evolve from point-like features at $\Gamma$ to enlarged pockets at deeper binding energy (Fig. 1(c)). This is corroborated by dispersions measured along the $\Gamma$-M-K-$\Gamma$ direction. Along this high symmetry direction, a series of highly dispersive valence bands are centered at the $\Gamma$ point and merge into relatively flat dispersions in the energy range between -1.0 and -1.5 eV. From comparison to orbital-projected DFT calculations, the highly dispersive bands near the VBT in the FM state are dominated by the Te 5$p$ orbital (Fig. 1(f)), while the Cr 3$d$ orbitals are mostly concentrated within the energy range between -1.0 eV to -1.5 eV~\cite{DFT_method}. The overall electronic structure below \tc~as shown in Fig. 1(d) is in qualitative agreement with the DFT calculations (Fig. 1(e) and Fig. 1(f)). 


\begin{figure}
\includegraphics[width=0.5\textwidth]{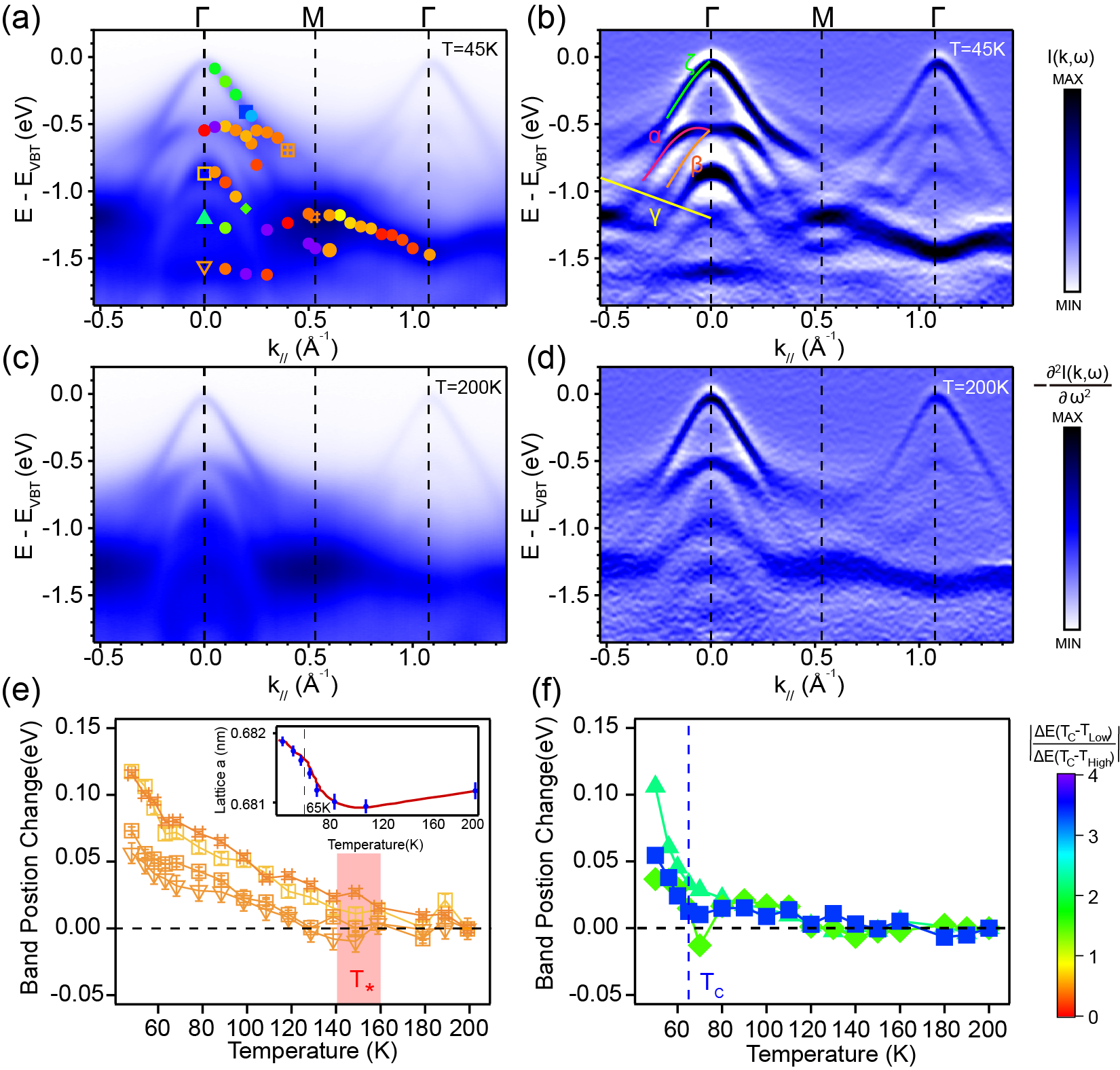}
\caption{Temperature evolution and analysis.
(a) Raw spectral image and (b) its second energy derivative along $\Gamma$-M-$\Gamma$ measured at 45K (T $<$ T$_C$ = 65 K). (c)-(d) Same as (a)-(b) but measured at 200K. (e)-(f) Fitted band position as a function of temperature taken at energy/momentum points as labeled by unique markers in (a). The color of each fitted marker represents the abruptness of change at T$_C$, defined as ~$\left |\frac{\Delta E_{T_c-T_\text{low}}}{\Delta E_{T_c-T_\text{high}}}  \right |$~. All fitted points are shown in (a), while those with a temperature onset (T$_*$) well above T$_C$ are shown in (e) and those with changes at T$_C$ shown in (f). The inset in (e) shows the lattice $a$ as a function of temperature adapted from Ref.~\cite{Carteaux1995}. The blue and red lines in (e) and (f) mark the two different temperature scales.
}\label{fig:Fig2}
\end{figure}

To pave the way for understanding the temperature-induced evolution of the electronic structure, we present temperature-dependent ARPES data in Fig. 2. From dispersions along the high-symmetry $\Gamma$-M-$\Gamma$ direction measured at 45 K and 200 K, we observe that the insulating nature is persistent across \tc~(Fig. 2(a)-(d)), namely that an electronic gap remains (see Supplemental Material~\cite{SM} for the gap size discussion). This is in disagreement with the non-magnetic DFT calculation (Fig. 1(g)), which predicts a metallic state. As the Cr $t_{2g}$ states are partially filled, this inconsistency suggests that local moments likely survive well above \tc~into the paramagnetic phase, driving the system into a Mott insulating state, consistent with the previous report on its sister compound Cr$_2$Si$_2$Te$_6$~\cite{williams2015}. There are, nevertheless, observable changes across the temperature range. The curvature of the $\alpha$ and $\beta$ band tops, marked in pink and orange in Fig. 2(b), changes with temperature. In addtion, the linear-like band, labeled as $\gamma$, disappears with temperature. 
Besides the overall changes of the $\alpha$, $\beta$ and $\gamma$ bands, all bands shift with temperature. To better understand the evolution of the bands, we perform a detailed analysis of the energy distribution curves (EDCs) across the $\Gamma$-M-$\Gamma$ cut by fitting and tracking the location of the observable bands (see Supplemental Material (SM) at ARPES Measurements section and Fig. S1-3 for more details~\cite{SM}). All fitted bands and momentum points are labeled by unique markers in Fig.~\ref{fig:Fig2}(a). Interestingly, for all the fittable bands, we can identify two temperature scales where shifts in the band position onset, one at \tc~and the other (T$_*$) around 150 K. 
To better visualize each band's tendency to shift at the two onset temperatures, we take the ratio of the band shift between \tc~and 200 K and between 50 K and \tc, ~$\left |\frac{\Delta E_{T_c-T_\text{low}}}{\Delta E_{T_c-T_\text{high}}}  \right |$~ as the color scale for each marker on each point. A larger value indicates a greater change of the band position at \tc~while a value smaller than 1 indicates a larger change at the higher temperature scale T$_*$. As a result, the changes that are strongly correlated with \tc~have a cold color on this scale reflecting a value above 2, and changes correlated with the higher temperature scale will have a warm color reflecting a value below 1. Interestingly, most bands shift gradually across T$_C$ except those near the VBT marked by green lines named $\zeta$ band in Fig.~\ref{fig:Fig1}(b), where the changes occur primarily near \tc. This is strongly correlated with the orbital character of the bands, with those primarily associated with Cr 3$d$ smoothly evolving across \tc~and those with Te 5$p$ shifting abruptly across \tc. To better demonstrate this distinct temperature behavior, we plot the temperature-dependent shift for those with an onset primarily at T$_*$ in Fig.~\ref{fig:Fig2}(e), and those with only an onset primarily at \tc~in Fig.~\ref{fig:Fig2}(f), all referenced to the final band position at 200 K. The dichotomy of the temperature behaviors is clearly contrasted. 

As there are no known phase transitions above \tc, the high temperature scale in the band evolution is likely associated with the response of the electronic structure to fluctuation effects associated with the FM order. 
In Cr$_2$Ge$_2$Te$_6$, it has been reported that the lattice parameter $a$ (inset in Fig. 2(e)) shows a negative thermal expansion with an onset temperature near 100 K, well above \tc~\cite{Carteaux1995}. To investigate the direct impact of a temperature-dependent lattice change on the electronic structure, 
we performed DFT calculations using the temperature-dependent lattice parameters refined at 5 K, 70 K and 150 K by neutron scattering experiment (see Fig. S6~\cite{SM} in the Supplemental Material for additional neutron scattering data)~\cite{Huang2021-my3}. However, both the direction and magnitude of band shifts are not consistent with the observed band shifts, suggesting that the changes in the electronic structure, particularly those that set in at high temperatures, cannot be directly accounted for the temperature-induced lattice change. 


\begin{figure}
\includegraphics[width=0.5\textwidth]{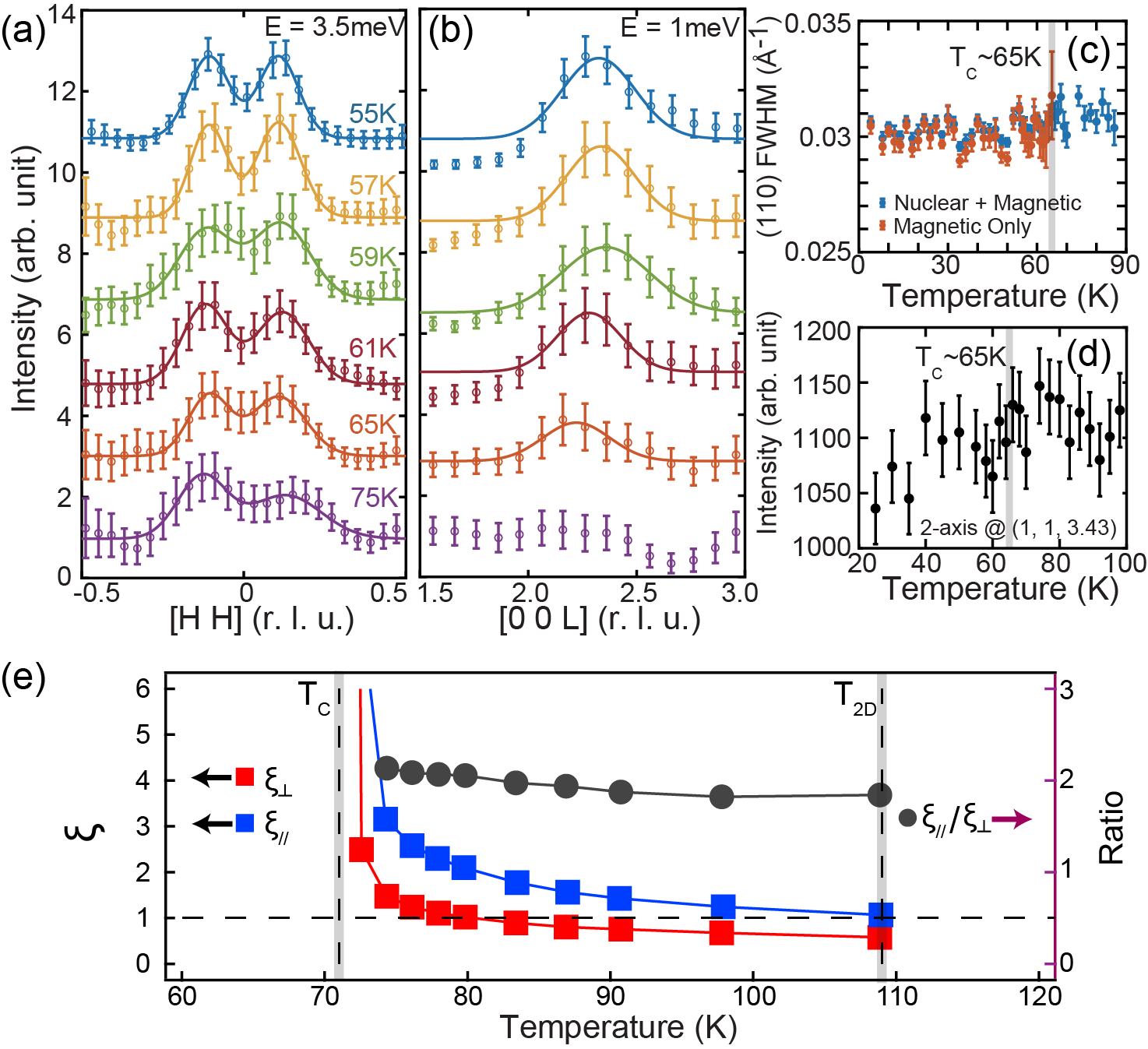}
\caption{Neutron scattering and Monte Carlo simulations. 
(a)-(b) Magnetic excitations measured via inelastic neutron scattering along [H H 0] and [0 0 L] as a function of temperatures across T$_C$, the solid lines are Gaussian fits. (c) The elastic peak width of the (110) peak measured by neutrons; (d) The two-axis measurement at wavevector (1, 1, 3.43). (e) The simulated change in the in-plane $\xi_\parallel$ and out-of-plane $\xi_\perp$ correlation lengths, in units of the nearest-neighbor lattice spacing, as well as their ratio, as a function of temperature. The crossover scale $T_{2D} \sim 2T_c$ when $\xi_\parallel=1$. The horizontal dotted line is a guide to the eye denoting the nearest-neighbor spacing with T$_{2D}$ indicated here as the point where $\xi_\parallel$ begins to exceed this distance. 
}\label{fig:Fig3}
\end{figure}

To gain insights into the possible origin of the higher temperature band evolution, we consider that for a quasi-2D magnet, while the in-plane and out-of-plane spin correlations both diverge at the same rate near \tc, the in-plane correlations, due to larger in-plane exchange interactions, would exceed several lattice constants at higher temperatures compared to that of the out-of-plane correlations, manifesting in a 2D to 3D crossover behavior, as has been reported for the iron-based superconductors~\cite{Wilson2010,Yi2019-my1,Huang2022-my2}. In order to investigate this behavior, we probed the magnetic correlations by neutron scattering. 
First, we observe no broadening of the (110) nuclear and magnetic elastic peaks around \tc~(Fig.~\ref{fig:Fig3}(c)). Second, we performed a two-axis experiment at (H,K,L)=(1, 1, 3.43). The wave-vector-dependent magnetic susceptibility is proportional to the signal integrated over the energy transfer E=$\frac{{k_i}^2 - {k_f}^2}{2m_n}$, with m$_n$ being the mass of neutron~\cite{chen2020}. This integration is achieved with triple-axis spectrometer by removing the analyzer to accept all neutrons along the final wave vector $k_f$, which was set $||$ c. If the magnetic fluctuation is 2D in the ab-plane, the magnetic scattering forms a ridge along the [1 1 L] direction. Only at a particular L value, 3.43 in this case, can the condition for the integration over the magnetic ridge be met. Therefore, the two-axis experiment can probe instantaneous spin correlations. No critical 2D divergence in the instantaneous correlations is observed at this wavevector across \tc, as expected since ultimately this is a 3D phase transition (Fig. 3(d)). Interestingly, previous experiments report a set of critical exponents consistent with those near a 3D tricritical point--a second order to first order crossover point~\cite{Chen2022,Lin2017,ME_1967}. 
To further determine the temperature dependent spin excitations across \tc, we performed inelastic neutron scattering experiments~\cite{Granroth_2010,Stone2014}, where we measured the spin excitation spectrum at different temperatures around \tc. From Fig. 3(a) we see that the in-plane spin excitations do not vanish up to 75 K and independent measurements show that in-plane spin excitations persist up to at least 150 K (see Fig. S4 in SM~\cite{SM} and more discussion about temperature dependence of the neutron scattering data in Supplemental Material), indicating the existence of short-range spin-spin correlations above \tc. For comparison, we measured the out-of-plane spin excitations (Fig. 3(b)), which shows a diffusive pattern instead of well-defined spin excitations at 75K, confirming that the out-of-plane spin correlation length drops below a lattice constant at temperatures slightly above \tc~while the in-plane short-range order persists to temperatures well above \tc. This behavior is 
consistent with the expected development of spin correlations in quasi-2D magnets.


To substantiate this understanding, we carried out classical Monte Carlo simulations using a classical Heisenberg model with the anisotropic exchange couplings from previous reports~\cite{Gong2017,MC_method,pawig_monte_1998}. As shown in Fig. 3(e), while both the in-plane correlation length $\xi_\parallel$ and the out-of-plane correlation length $\xi_\perp$ diverge at \tc, where the ratio of $\xi_\parallel$/$\xi_\perp$ is constant, $\xi_\parallel$ grows beyond the nearest-neighbor spacing ($\xi>1$) at a temperature scale ($T_{2D}$) much higher than that for the out-of-plane correlation. Note that the $T_{2D}$ is not identical to T$_*$, T$_*$ is experimentally determined and arises from the anisotropy of the in-plane and out-of-plane exchange coupling. Hence we can understand that in the range T$_C$$\textless$T$\textless$T$_{2D}$, 2D regions of short-range correlated magnetic moments begin to form in-plane while the different planes essentially remain uncorrelated, sustaining in-plane spin waves. The electronic structure responds in turn by the shift of Cr d-dominated bands. Just above \tc, out-of-plane correlation length reaches the lattice constant, leading to rapid response in Te p-dominated bands. 

\begin{figure*}
\includegraphics[width=0.95\textwidth]{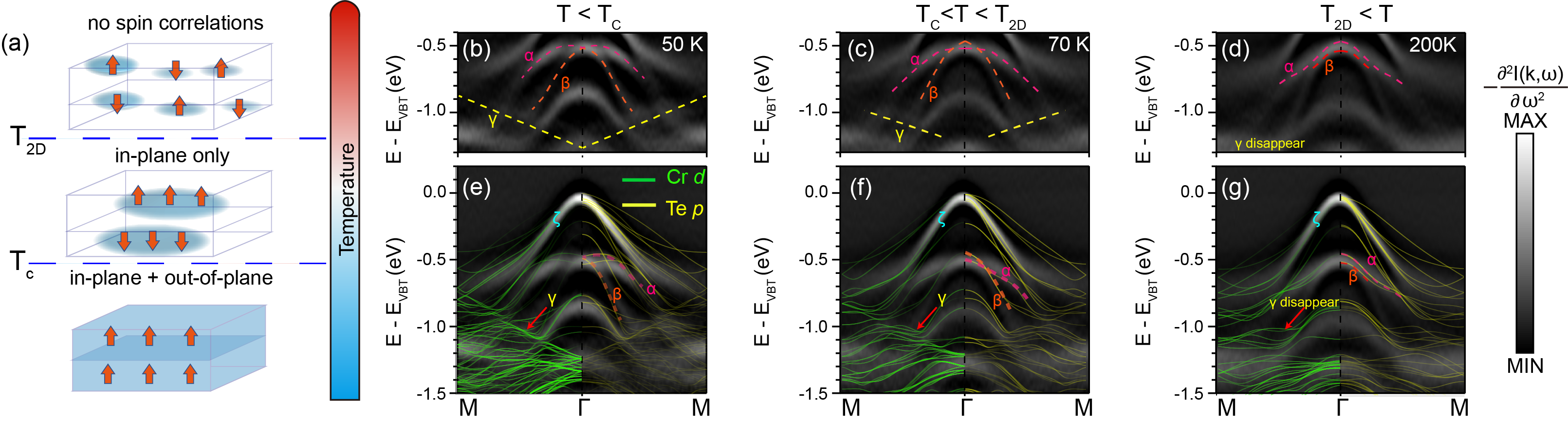}
\caption{Spin correlation model and DFT calculations.
(a) Schematic for the temperature evolution of spin correlations. Starting with no spin correlations at high temperatures, the in-plane correlations length exceeds the lattice constant first, followed by out-of-plane correlations length, leading to long range FM order at T$_C$. (b)-(d) Second energy derivatives of ARPES spectra from the three temperature regimes, with eye guides for dominant spectral change. (e)-(g) Same spectra images with corresponding DFT calculations for FM, A-type AFM and G-type AFM to simulate the three regimes, respectively. The green and yellow lines represent the band structures contributed by Cr 3d and Te 5p projected orbitals, respectively.
}\label{fig:Fig4}
\end{figure*}

With this understanding, we finally discuss the implications for the sequential band evolution observed by ARPES in comparison to DFT calculations. To model the development of in-plane and out-of-plane spin correlations, we model the highest temperature phase above $T_{2D}$ with all single spins anti-aligned,
which effectively mimics the persistence of local moments with no ferromagnetic spin correlations. For $T_{2D}>T>T_C$, we simulate the in-plane FM spin correlations via A-type AFM spin arrangement, where in-plane spins are co-aligned. The lowest temperature $T<T_C$ is the FM order where all spins are aligned along the $c$ direction.
The orbital projected DFT calculations for these three different models are plotted on top of the corresponding ARPES data (Fig. 4). For the lowest temperature phase ($T<T_C$), the dip-like shape for the $\alpha$ and $\beta$ band top and the linear $\gamma$ branch in ARPES data are all reproduced by the DFT calculations (Fig. 4(b),(e)). As the temperature rises above \tc~($T_{2D}>T>T_C$), the $\alpha$ and $\beta$ band top evolve into two hole-like bands centered at $\Gamma$. The $\beta$ branch with a larger slope and higher energy band top crosses the other branch, consistent with the DFT calculations. When the temperature goes up to 200 K ($T>T_{2D}$), the $\gamma$ band disappears, again consistent with the disappearance of this feature in the DFT model. As shown in Fig. 4(b) - (d), besides the great changes in the $\alpha$, $\beta$ and the $\gamma$ bands, the $\zeta$ band owns a smaller slope below \tc~and evolves into a larger slope dispersion around \tc~and exhibits no further dramatic changes above \tc. The $\zeta$ band's slope is also well captured by the Te $p$ orbital projected DFT calculations (yellow lines from DFT), suggesting the accuracy of the abstract model. The consistency between the DFT calculations for the three models and the measured temperature evolution of the band dispersions indirectly confirms our understanding that the band evolution manifests the response of the electronic band structure to the sequential development of spin correlations from 2D to 3D.

Overall, supported by our combination of ARPES, neutron scattering, DFT calculations and Monte Carlo simulations, we come to a comprehensive understanding of the development of the FM order in \cgt. Local moments from Cr $d$ orbitals appear at very high temperatures, resulting in an electronically gapped system. Due to the much larger exchange coupling in the in-plane direction, the correlation length in the in-plane direction exceeds that of a characteristic length scale first, causing bands with predominantly Cr 3$d$ character to start evolving well above \tc. With further lowering of the temperature, the out-of-plane correlation length reaches the characteristic length scale near \tc, and is manifested most strongly in the bands associated with Te $p$ orbitals, which play a critical role in bridging the magnetism along the c-direction via the Cr-Te-Cr superexchange interactions.
The development of spin correlations is typically probed by neutron scattering, which is limited to bulk crystals and not applicable for exfoliated vdW flakes.
As a vdW system with a single untangled magnetic order, \cgt~allows us to clearly demonstrate the sensitivity of using orbital-dependent electronic structure to track and resolve the development of spin correlations through the dimensional crossover of the magnetic order in quasi-2D magnets. We anticipate that this sensitivity can be potentially useful in probing the development of spin correlations in the few layer or even monolayer regime, and contribute to the understanding of low dimensional magnetism in the wider class of vdW magnets with more complex order parameters.

\section{acknowledgments}

This research used resources of the Advanced Light Source, and the Stanford Synchrotron Radiation Lightsource, both U.S. Department Of Energy (DOE) Office of Science User Facilities under contract nos. DE-AC02-05CH11231 and AC02-76SF00515, respectively. ARPES work at Rice is supported by the U.S. DOE grant No. DE-SC0021421, the Gordon and Betty Moore Foundation’s EPiQS Initiative through grant no. GBMF9470, and the Robert A. Welch Foundation grant no. C-2175 (M.Y.). The neutron scattering and single crystal synthesis work at Rice was supported by US NSF-DMR-2100741 and by the Robert A. Welch Foundation under Grant No. C-1839, respectively (P.D.). 
Work at University of California, Berkeley, is funded by the U.S. Department of Energy, Office of Science, Office of Basic Energy Sciences, Materials Sciences and Engineering Division under Contract No. DE-AC02-05-CH11231 (Quantum Materials program KC2202). 
Work at Los Alamos was carried out under the auspices of the U.S. Department of Energy (DOE) National Nuclear Security Administration (NNSA) under Contract No. 89233218CNA000001, and was supported by LANL LDRD Program and in part by Center for Integrated Nanotechnologies, a DOE BES user facility, in partnership with the LANL Institutional Computing Program for computational resources.
A portion of this research used resources at the High Flux Isotope Reactor and Spallation Neutron Source, a DOE Office of Science User Facility operated by the Oak Ridge National Laboratory.
M.W.B. was funded by the Robert A. Welch Foundation grant no. C-1818. A.H.N. acknowledges the support of the National Science Foundation grant no. DMR-1917511.

\bibliography{bib_correctedstyle.bib}

\end{document}